\documentclass[aps,twocolumn,a4paper]{revtex4}

\usepackage{graphicx}

\begin{document}



\title{Rotation-induced 3D vorticity in 
$^4$He superfluid films adsorbed on a porous glass }
\author{M. Fukuda$^1$,  M.K. Zalalutdinov$^1$,  V. Kovacik$^1$,  T.
Minoguchi$^2$, T. Obata$^1$, M. Kubota$^1$, and E.B. Sonin$^{1,3}$  }

\affiliation{$^1$Institute for Solid State Physics, University of Tokyo,
Kashiwa, Chiba 277-8581, Japan
\\ $^2$ 	Institute of Physics, University of Tokyo, Komaba,
Meguro-ku, Tokyo 153-8902, Japan
\\\ $^3$Racah Institute of Physics,  Hebrew University of Jerusalem, Givat Ram,
Jerusalem 91904, Israel }

\date{\today}

\begin{abstract}

Detailed  study of torsional oscillator experiments under steady rotation up to 6.28 rad/sec is reported for a $^4$He superfluid monolayer  film formed in 1 $\mu$m-pore diameter porous glass. We found a new dissipation peak with the height being in proportion to the rotation speed, which is located to the lower temperature than the vortex pair unbinding peak observed in the static state. We propose that 3D coreless vortices ("pore vortices") appear under rotation to explain this new peak. That is, the new peak originates from dissipation close to the pore vortex lines, where large superfluid velocity shifts the vortex pair unbinding dissipation to lower temperature. This explanation is confirmed by observation of nonlinear effects at high oscillation amplitudes. 

\end{abstract}


%

\maketitle
Superfluid $^4$He films adsorbed in porous media provide unique 
possibility to study the interplay between 2D and 3D physics\cite{Rp}. 
Especially for the superfluid transition, the 
system shows similar behavior to the 2D film (the 
Kosterlitz-Thouless transition\cite{KT}) such as the 
density proportional superfluid transition temperatute $T_c$, with the energy 
dissipaton peak around  $T_c$ \cite{ Rp,  Shir, KW, Gabor, Mikhin}. Whereas the superfluid density $\rho_s$ critical  index is found close to $2/3$ \cite{Rp} of 3D system and a sharp cusp of 
the specific heat\cite{Murphy} appearing around $T_c$, which is 
similar to the $\lambda$ transition of the bulk $^4$He. 

A crucial role in the Kosterlitz -Thouless 2D transition belongs 
to thermally excited vortex-antivortex pairs (VAPs). 
Also for the $\lambda$ transition 
of the 3D system, a similar mechanism is proposed where the 
vortex rings play the important role\cite{GW}.  
However, multiple connectivity of the superfluid film in porous media allows a variety of vortex configurations other than VAP, e.g., vortex rings and 3D coreless pore vortices \cite{Min, Mahta, GW, OK}.  Therefore one may expect that the response of the film on a porous substrate would be essentially different from that on a plane substrate, and observation under rotation
should be quite interesting for understanding the role of these vortices on the superfluid transition.


An effective experimental method to study superfluidity of films, both on a plane
and on a porous substrate, is the torsional oscillator technique (TO)\cite{Rpflt}.
The vortex dynamics of 2D films, which is probed in torsional oscillator
experiments, was developed by Ambegaokar, Halperin, Nelson and Siggia (AHNS)\cite{AHNS}. 
The theory and the experiment revealed that close to the critical temperature, the VAPs give a rise to a dissipation peak, where VAPs start to dissociate and free vortices appear according
to the AHNS theory\cite{AHNS}. 
The vortices in the
superfluid film are created by thermal fluctuations and therefore their number is strongly
temperature-dependent and also depends on parameters which are difficult to
measure or estimate independently. On the other hand, rotation with the angular
velocity $\Omega$ produces free vortices with temperature-independent areal density
$2\Omega/\kappa$ ($\kappa=h/m$ is $^4$He circulation quantum), and this
makes the dynamical properties easier for the analysis. Adams and Glaberson\cite{AdGl} have shown experimentally that rotation-induced vortices in the
plane superfluid film 
responsible for essential dissipation at low
temperature side of dissipation, where the density of thermally created vortices is negligible and
dissipation is not detectable without rotation. They obtained valuable information on 2D vortex diffusivity from the low-temperature dissipation tail under the rotation.  


We investigate  monolayer superfluid films covering  a 1$\mu$m-pore porous-glass substrate\cite{Kubota} under rotation by the TO technique. We have chosen the porous glass with this pore diameter because of the clear thin film feature of He and apparent 3D connectivity. Namely He atom de Broglie wave length is much smaller than the pore diameter under the conditions of present study and our situation is closely connected to the one considered by theoretical works for thin films\cite{Min, Mahta, OK}.  
The 2D superfluid density (per unit area) $\rho_{s2}$ is derived from the period change of the oscillation, while the information about the vortex dynamics is obtained from the dissipation, characterized by the change $\Delta Q ^{-1}$ of the quality factor $Q$. Traditionally the response of the superfluid film in the TO experiment is described by the ``dielectric permeability''
$\varepsilon$, which is the ratio of the effective mass of the superfluid component participating in the oscillation to the total, or ``bare'' superfluid mass. Then
$\Delta Q^{-1}=(M_s/M_{TO})\mbox{Im} \left( {\varepsilon ^{-1}} \right)$,
where M$_s=A \rho_{s2}$ is the total superfluid mass of the film of area $A$
and M$_{TO}$ is the oscillator's effective mass.

The design and performance
of the rotating dilution refrigerator used were reported previously\cite{rotref}. 
It can be rotated with the angular velocity up to 6.28 rad/sec,
while temperature of the mixing chamber is controlled in the range, 50 mK-2 K. 
A magnetic seal unit connects the cryostat with the stationary gas handling
system and provides continuous operation during rotation. Be-Cu TO (f=477Hz,
quality factor Q$_{100mK}$=10$^6$) with the stack of the porous glass disks
(diameter $\Phi $=15~mm, total height 9~mm)
is mounted  
below the mixing chamber through a vibrational isolation scheme.

\begin{figure}
\begin{center}
    \leavevmode
    \includegraphics[width=1.0\linewidth]{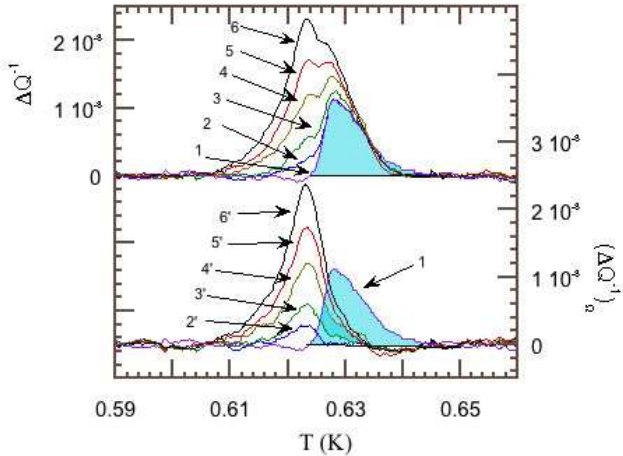}
    \bigskip
\caption{ Energy dissipation peak under rotation (a), the 
number from 1 to 6 each corresponds to  rotation speed 0, 0.79,
1.57, 3.14, 4.71, 6.28~rad/$\sec$ respectively.  It is realized that the data in (a) is actually the summation of the two peaks, namely the static peak and and  a rotationally induced peak, by subtracting the static peak 1 in (b) from the data displayed in (a).   Rotationally induced peak $(\Delta $ Q $^{-1})_\Omega $is displayed in (b) together with the static peak with dashed number, each corresponding to the number in (a). Data are all for the film with the superfluid transition temperature T$_c$=628mK, which is determined as in \cite{Shir}. See also Fig. 2 caption.}
\label{Fig.1} \end{center}
\end{figure}

After studying superfluid density (period shift) and dissipation (Q value)
from the lowest temperature to above $T_c$,
TO is cooled down to 0.95$T_c$, where $T_c$ is the superfluid onset temperature determined in the same manner as in \cite{Shir},
then rotation of the cryostat at angular velocity 0 $\leq \Omega \leq
$ 6.28rad/sec is started and period and amplitude of the TO are recorded while
the temperature is slowly swept up to T=1.05 $T_c$  during 20 hours for each $\Omega$.
This procedure provides reproducibility of period and amplitude data of the TO. In
order to calculate  $\Delta Q ^{-1}$ the level of the background is
measured before this measurement cycle and is subtracted from the data,
which are shown in Fig.\ \ref{Fig.1}.
Each curve in the upper frame represents $\Delta Q ^{-1}$ 
near the superfluid transition under rotation
with the angular velocities from 0.79 to 6.28 rad/sec.

\begin{figure}
\begin{center}
    \leavevmode
    \includegraphics[width=0.85\linewidth]{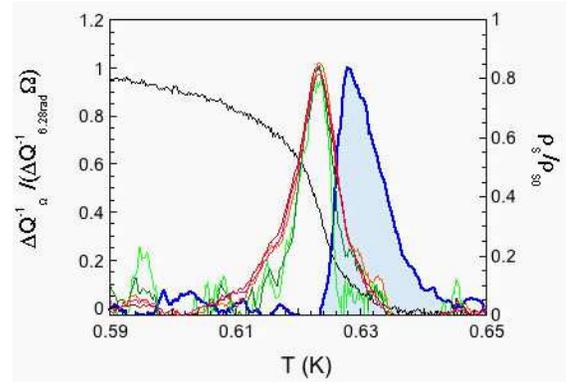}
    \bigskip
\caption{ The right side (high T) peak is  the one appears in static
condition. The group of the left peaks is rotation-induced ones
for various $\Omega$ scaled (divided) by $\Omega$. The fact that all the curves
collapse more or less on the same curve proves linearity on $\Omega$. In addition, superfluid density scaled by its zero temperature value is displayed. The lagest slope is extraporated to zero density to T$_c$=628mK where the static peak is also located\cite{Shir}.}  
\label{Fig.2} \end{center}
\end{figure}

The location of the static peak  in comparison to the superfluid density is 
different from flat film case\cite{Shir}. Namely it is located at higher temperature than rapid change in superfluid density occurrs.
The period shift curves are found to be common for the data in static  
state and under rotation, i.e., there was no detectable change in $\rho _s$ under
rotation in this angular velocity range (Fig.\ref{Fig.2}). 
Under rotation two
pronounced features of dissipation  were detected, which are dramatically different 
from the response of flat films to rotation:
\begin{enumerate}

\item There was no low-temperature tail of dissipation, which characterized the
response of the flat film to rotation \cite{AdGl}. Instead a sharp cut-off of 
dissipation at the low temperature side of the peak was observed.

\item  The double-peak structure of the
dissipation was revealed with sharp peaks at constant temperatures.

\end{enumerate}

A peak, which is seen on  the right (high T) side of rapid change of superfluid density in Fig. \ref{Fig.2}, is present even in the
static state, $\Omega$=0. The other (left) one, which is found only under rotation, grows
proportionally to the rotation speed,$\Omega$. And the peak is located at a constant temperature
about $1.5 \%$ below $T_c$, where 
  no excess dissipation was detected when $\Omega$=0.
In order to demonstrate that rotation really produces a new dissipation peak $\propto
\Omega$, Fig.\ \ref{Fig.2} shows the rotation-induced dissipation peaks
$\Delta Q^{-1} _{\Omega }=\Delta Q^{-1} -\Delta Q^{-1} _s$, where the
contribution
$\Delta Q^{-1} _s$ of the static state was subtracted from the total dissipation,
as a function of temperature. Then rotation induced peaks $\Delta Q^{-1}
_{\Omega }$ were
scaled (divided) by $\Omega$ and $\Omega_{max}$=6.28 rad/sec. They collapse on the same curve except for low temperature ends, which are worth of  more careful analysis. The
half width of the rotational peak is about
$\Delta T/T_c \approx$1\%.

\begin{figure}
\begin{center}
    \leavevmode
    \includegraphics[width=0.92\linewidth]{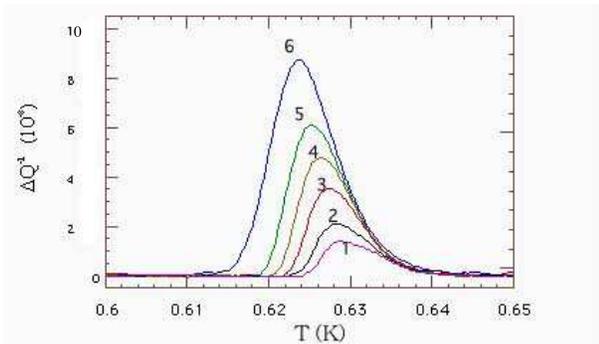}
    \bigskip
\caption{ Energy dissipation curves in static condition  for non-linear regime. 
 AC drive velocity 
for each curve corresponds to V$_{AC}$
=0.095(No.1), 0.19, 0.36, 0.52, 0.66, 0.94(No.6)~cm/ $\sec$ respectively. }
\label{Fig.3} \end{center}
\end{figure}

The amplitude of the torsional oscillation, used for the recording of the
data in Fig. \ref{Fig.1}, corresponds to the drive AC velocity of the TO V$
_{AC}$
=0.03~cm/sec,
which is
an average velocity calculated from the oscillator dimensions and it is in the 
linearly responding region. We have also measured dissipation peaks at a wide 
range of AC amplitude under static 
as well as under rotation conditions.  Fig. \ref{Fig.3} presents the peaks under static
 condition at the range of  drive AC velocity, 0.09~cm/$\sec
<$V$_{AC} <$ 0.9~cm/$\sec$.
The non-linear effect becomes pronounced above V$_{AC} \sim 0.25$ cm/sec.

\begin{figure}
  \begin{center}
    \leavevmode
    \includegraphics[width=0.92\linewidth]{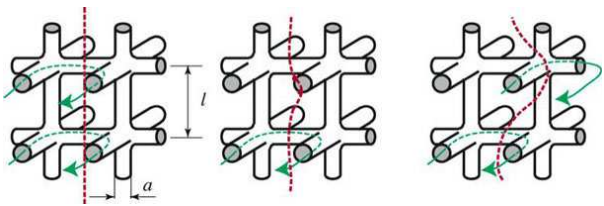}
    \bigskip
    \caption{Vortex creep in the jungle gym structure: the vortex (dashed) line
crosses a cylinder between two cells, creating a VAP in the cylinder. The
arrowed
curved lines show circulation around the cells where the vortex line is
located.}
\label{FigT}  \end{center}
\end{figure}

Now, we shall utilize these nonlinear response data in explaining the
rotational dissipation peak. We model the porous medium with the
``jungle gym'' structure\cite{Min}, which is a 3D cubic
lattice of intersecting cylinders of diameter $a$ with period $l$ (Fig. \ref{FigT}). If
this structure is a substrate for a superfluid film, multiple connectivity allows a new
type of vortices, which is impossible in a flat single connected film: a 3D vortex
without a 2D core, i.e., without any phase singularity in the film. The 
position of vortex line is confined within the pores around which there is a superfluid
velocity circulation. Vortex motion is possible as a creep process, when the vortex
line shifts from one pore to another by crossing  cylinders. Crossing of a cylinder means 
creation of a VAP in the film, which covers this cylinder, as shown in Fig. \ref{FigT}.

If the vortex line moves on distances much larger than the structure-cell size
$l$ one
may describe the vortex motion   by a continuous displacement of the vortex
line. But
for small-amplitude oscillations the position-vector has a poor physical
meaning,
because it can be defined only with accuracy of the structure period $l$.
In practice
this means that the vortex is locked at some chain of pores by
``intrinsic pinning'' (a concept suggested for vortices in layered
superconductors).
However, if the vortex line cannot move, the question arises, how it is possible to
detect its
presence at all. We argue that though the  3D vortex lines cannot move
themselves once 3D coherence is established, they can
influence  the gas of VAPs, and the rotation experiment reveals this
influence. This
immediately explains the first feature of the rotation response: no
low-temperature
dissipation. Indeed, an essential number of VAPs is possible only close to the
transition temperature. Since one can notice the rotation-induced vortex lines
only via
their effect on VAPs, one can reveal the presence of locked 3D coreless
vortices also
only at high temperatures where 2D VAPs are present.

In order to explain the second feature of the rotation response (the
double-peak structure), we shall
consider the effect of circular currents around the 3D vortices on the VAP
dissipation.
At the highest rotation speed  $\Omega = 6.28$rad/sec, the average inter-3D vortex
distance is
$L_v = \sqrt{\kappa/2\Omega} \simeq 90[\mu m]$. This means that the 3D
vortices  are well
separated: $L_v$ is  about  90 times larger than $\ell$ under rotations in
question. As
a result of it,  the 3D-vorticity generated superfluid velocity relative to
the substrate
is effectively zero except for the regions close to the 3D vortex centers.
For each 3D
vortex, the length of the squares of cylinders  surrounding the center is
$4(2n-1)
\ell$.  Here $n=1$ is of the closest square  to the center, $n=2$ is of the
2nd closest
one, etc. The superfluid velocity for the  circulation in the $n$-th
closest square is then
$v_s^{(n)} = 
\kappa /\{4(2n-1) \ell\}$. (As $n$ goes larger, the circulation
changes to circular from square. Here we neglect this change because it changes
$v_s^{(n)}$ only by a numerical factor.)
Here we  assume $\ell = 2.5 a$, which gives
reasonable aspect ratio seen from SEM\cite{Kubota} and STM  observations. For $a=1[\mu
m]$, we obtain
$v_s^{(1)}=1.0$[cm/s], $v_s^{(2)}=0.33$[cm/s], $v_s^{(3)}=0.2$[cm/s],
$v_s^{(4)}=0.14$[cm/s], $v_s^{(5)}=0.11$[cm/s], and so on.

\begin{figure}
 \begin{center}
    \leavevmode
    \includegraphics[width=0.7\linewidth]{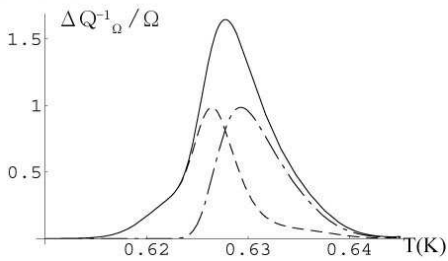}
    \bigskip
\caption{Calculated rotation-induced energy dissipation peak
$\Delta Q^{-1}_\Omega$ devided by $\Omega$ (dashed line)
along with the peak at the static state $\Delta 
Q^{-1}_s$ (dash-dotted line). To compare with Fig.2, 
these two peaks are scaled so that their peak heights 
become unity. The solid line is the simple add of them.}

\label{Fig.5} \end{center}
\end{figure}

Assuming that these circular flows around vortex lines produce the same
effect on
dissipation as large amplitude AC flows, we shall utilize corresponding
large amplitude AC dissipation data  in Fig. \ref{Fig.3} respectively at
$v_s^{(1)} = 0.94$[cm/s], $v_s^{(2)}=0.36$[cm/s], $v_s^{(3)}=0.19$[cm/s] and
$v_s^{(n)}  = 0.095$[cm/s] for $n \geq 4$: The rotational
dissipation $\Delta Q^{-1}_{\Omega}$ mainly comes from the nonlinear
dissipation of VAPs
in $v_s^{(n)}$ with $n \leq 3$ of the 3D vortices; 
\begin{eqnarray}
\Delta Q^{-1}_{\Omega}
&=& \frac{1}{S_{tot}}
         \sum_{n\geq 1} S_n (\Delta Q^{-1}_{(n)}-\Delta Q^{-1}_s) \nonumber \\
&=& C \sum_{n=1}^3 (2n-1) (\Delta Q^{-1}_{(n)}-\Delta Q^{-1}_s)~,  
\label{result}
\end{eqnarray}
where $\Delta Q^{-1}_{(n)}$ are data from Fig. \ref{Fig.3} for $v_s^{(n)}$,
$S_n \equiv 4(2n-1) \ell \pi a$ is the substrate surface area in the $n$-th
closest square and $S_{tot} \equiv 2 \pi a \ell \times L_v^2/\ell^2$ is
the total
substrate area per single vortex. 
The prefactor takes $C=4 \Omega \ell^2 / \kappa \simeq
1.6 \cdot 10^{-3} \Omega$ when $\Omega$ is measured by radian per second.
Then, the
height of $\Delta Q^{-1}_{\Omega}$ has turned to be
\begin{equation}
[\Delta Q^{-1}_{\Omega}]_{max} = 3.5 \cdot 10^{-10} \Omega
\end{equation}
with $\Omega$ in $[rad/s]$.
This value is  $10$ times smaller than the experimental one, although the
peak position at $0.626$[K] is in  good agreement with the experiment (a
little bit higher than the experimental one). In Fig. \ref{Fig.5}, the rotation-induced
dissipation  multiplied by this correction $10$ is plotted along with the
dissipation of the static  state, which looks similar to  the experimental one shown in
Fig.\ref{Fig.2}. The total dissipation should be a simple sum of these two
dissipations. But the double-peak  structure, which one can
see in the experimental curve in Fig. \ref{Fig.1}, would appear  in the
calculated total dissipation only if the calculated rotation dissipation peak were 
 narrower. Altogether, this means that our model based on using the
experimental data on large-amplitude AC dissipation yields the rotation 
dissipation peak in 10 times lower in height and 
somewhat narrower than the experimental one. This disagreement does
not look discouraging bearing in mind complicated geometry for AC and
DC external flows. One of the reasons why our model predicts a weaker
dissipation is because the large-amplitude AC flows change its direction 
every period and it should weaken their effect on dissipation comparing 
with static circular flows generated by 3D vorticity.


In conclusion, we demonstrated the existence of an
additional dissipation peak whose height changes proportionally to the rotation speed in the
torsional-oscillator experiment with submonolayer superfluid
$^4$He films in porous glass. We relate the rotation-induced
peak to 3D
coreless pore vortices 
threading through the jungle-gim strucure cells. We
argue that the rotation peak is due to high superfluid velocity around
centers of
3D vortices. This effect is similar to the nonlinear effect of high-amplitude
oscillating
velocity, which was revealed both for flat and multiplly-connected porous
substrates.  Using our own measurements of this nonlinear effect for our
system, we
were able to reproduce the structure of the observed rotation dissipation
peak within
a reasonable numerical agreement. One can expect larger seperation of
the peaks for
films condensed on smaller pore size porous glass and for larger pore
systems we would
expect merging of the two peaks into one, but it would be still different
from flat film responce. 

Authors express acknowledgments to W. I. Glaberson, J. D.
Reppy,  and W. F. Vinen for fruitful discussions. E.B.S.
acknowledges
hospitality and support of the Institute for Solid State Physics, the
Univ. of Tokyo. This work has been supported by Grant-in-Aid for Scientific Research from JSPS.


\end{document}